\documentstyle[epsf,aps]{revtex}

\begin{document}
\draft
\title {Coherent Control of Magneto-optical Rotation in\\
Inhomogeneously Broadened Medium}

\author {
Anil K. Patnaik\footnote{Present Address: Department of Applied Physics and Chemistry,
University of Electro-Communications, Chofu, Tokyo 182-8585, Japan} and G. S. Agarwal}
\address {Physical Research Laboratory, Navrangpura, Ahmedabad-380 009,
India}
\date{\today}

\maketitle
\begin{abstract}
We extend our earlier investigations [Opt. Commun. {\bf 179}, 97
(2000)] on the enhancement of magneto-optical rotation (MOR) to
include inhomogeneous broadening. We introduce a control field
that counter-propagates with respect to the probe field. We derive
analytical results for the susceptibilities corresponding to the
two circular polarization components of the probe field. From the
analytical results we identify and numerically demonstrate the
region of parameters where significantly large magneto-optical
rotation (MOR) can be obtained. From the numerical results we {\em
isolate} the significance of the magnetic field and the control
field in enhancement of MOR. The control field opens up many new
regions of the frequencies of the probe where large
magneto-optical rotation occurs. We also report that a large
enhancement of MOR can be obtained by operating the probe and
control field in two-photon resonance condition.
\end{abstract}

\section{Introduction}
A magnetic field, when applied to an initially isotropic medium containing
gaseous atoms having $m$-degenerate sublevels, can cause birefringence in
the medium. Because, the applied magnetic
field creates asymmetry between the susceptibilities $\chi_\pm$ of the
medium corresponding to the two circularly polarized components $\sigma_\mp$
of the probe field. That results in magneto-optical rotation (MOR), i.e.
the plane of polarization of a weak probe field is rotated when it passes
through the medium. For a small absorption the rotation angle $\theta$
is given by
\begin{equation}
\theta = \pi k_p l (\chi_- - \chi_+)~;
\end{equation}
where $\vec{k}_p$
corresponds to propagation vector of the probe and $l$ is length of the
cell along $\vec{k}_p$. Further we note that, $\chi_\pm$ depend on the
atomic density and the oscillator strength of the atomic transition.

    Traditionally MOR was used as a tool in polarization spectroscopy
using continuum sources \cite{kliger_book}. The interest in MOR was
intensified in the atomic and molecular physics with the availability of
intense light sources of definite polarization \cite{MOR_synch} and
frequency \cite{MOR_laser}. Several reviews exist in the literature
on this subject including several interesting applications (e.g., see
\cite{connerade_book}). Using saturating fields the non-linear
MOR has also been studied at length
\cite{MOR_sat2,MOR_sat1,gsa-sat,gawlik_review}.
MOR with a transverse magnetic fields \cite{voigt} (known as Voigt effect)
and with inclined magnetic fields \cite{MOR_inclined} have also been studied.
Recently large MOR has been reported in dense and cold atomic cloud of Rubidium
\cite{cold}. On the other hand, laser field {\em alone} can also
break the symmetry in the
response of an atomic gas to different polarization components of a probe
field. For example, let us consider $j=0 \leftrightarrow j=1$ transitions of
an atomic gas
containing $V$ systems. When a linearly polarized weak probe field passes
through the medium, the $\sigma_\pm$ components of the probe field couple
the $|j=0,m=0\rangle$ with the degenerate states $|j=1,m=\mp 1\rangle$. The
susceptibilities $\chi_\pm$
of the medium to these two components $\sigma_\mp$ are same; i.e., response
of the medium is symmetric to both the components. However when a $\sigma_-$
polarized strong field is applied on the $|j=0,m=0\rangle
\leftrightarrow
|j=1,m=1\rangle$ transition, the susceptibility $\chi_+$ is modified by the
control field parameters creating asymmetry between $\chi_+$ and $\chi_-$.
Thus the plane of polarization of the probe field rotates (due to Eq.\ (1)).
Note that, this rotation is solely due to
the laser field and is a function of control field parameters.
Resonant birefringence due to optically induced level shift
by a coherent source was observed \cite{biref_Lshift}.
The light-induced polarization rotation in optical pumping experiments with
incoherent light has also been extensively studied \cite{biref_incoher}.
Liao and Bjorklund \cite{liao:76} were the first to observe polarization
rotation in a three level system by resonant enhancement of two-photon
dispersion in the $3s^2 S_{1/2}\leftrightarrow 5s^2S_{1/2}$ of sodium vapor.
H\"{a}nch and coworkers
\cite{hansch:76} have used this polarization rotation as a high resolution
spectroscopic technique. Heller {\em et al} \cite{heller:81} extended this
idea to atomic systems involving the ionization continuum. Experimental
and theoretical work has been reported by St{\aa}hlberg {\em et al}
\cite{stahlberg:90} on laser induced dispersion in a three level cascade
system of $Ne$ discharge.

Recently, combining the ideas of enhancement of refractive index
using atomic coherence \cite{RI_scully} and the non-linear MOR,
Scully and his coworkers have investigated a possible application
to high-precision optical magnetometry
\cite{magneto_theory,magneto_expt}. They have demonstrated this
possibility both theoretically \cite{magneto_theory} and
experimentally \cite{magneto_expt}, considering the rotation of
polarization of a strong linearly polarized probe caused by an
optically thick cell containing $^{87}$Rb vapor. The maximum
sensitivity reported in their experiment is $\sim 6\times 10^{-12}
{\rm Gauss/\sqrt{Hz}}$, which is superior to other existing high
precision magnetometers. Budker {\em et al} at Berkeley have also
reported high sensitive optical magnetometry in a series of papers
\cite{magneto_budker}, based on the non-linear MOR involving
ultra-narrow resonances ($\simeq 2\pi \times 1.3$Hz) using special
cell with high quality anti-relaxation paraffin coating that
enables the atomic coherence to survive even after a large number
of collisions with the wall. Using a similar configuration, Budker
{\em et al} have shown reduction of the group velocity of light to
$\simeq 8~{\rm m/sec}$ in a non-linear magneto-optical system
\cite{budker_slow}. Further, Pavon {\em et al} \cite{pavon:97}
introduced the idea of {\em coherent control} to obtain
significant atomic birefringence in presence of
electromagnetically induced transparency (EIT) \cite{harris}.
Winelandy and Gaeta \cite{gaeta:98} used quantum coherence to
control the polarization state of a probe field. They reported a
large birefringence and hence a large polarization rotation in a
three-level cascade of $^{85}$Rb (see also
\cite{my_MOR,my_DOP_springer}). Using a similar configuration,
Fortson and coworkers \cite{forston:98} have showed a possible
utility of the polarization rotation at EIT to measure the atomic
parity non-conservation signal with a better efficiency. A
detailed discussion on the role of degenerate sublevels and effect
of the polarized fields on EIT has been discussed in
\cite{fulton}.

However, it is interesting to investigate the combined effects of
the laser field and the magnetic field in the context of {\em
coherent control} of the rotation of polarization. In our earlier
work \cite{my_MOR}, we have reported laser field induced
enhancement of MOR in {\em cold} atoms. In the present paper we
generalize the above work \cite{my_MOR} by including the thermal
motions of the atoms inside the cell (see Fig.\ 1). Here a large
broadening is introduced in the rotation signal. This could be
desirable to get large rotations for a broad range of probe
frequencies in presence of a control field. But on the other hand,
broadening reduces the magnitude of rotation considerably.
However, one can work with a denser medium when Doppler effect is
included in the calculation. Moreover, we have included {\em all
spontaneous decay events} involved in the $j=0 \rightarrow j = 1
\rightarrow j =0$ transitions of the system (unlike in
\cite{my_MOR}). Further we discuss a special case when the weak
probe field and strong control field are in two-photon resonance
with $|e\rangle \leftrightarrow |g\rangle$ transition, that gives
rise to large enhancement of MOR.

The organization of the paper is the following. In Sec. II, we describe the
model scheme and determine the susceptibility of a moving atom using density
matrix formalism. In Sec. III, we present the analytical results for the
susceptibilities
of the Doppler broadened medium. In Sec. IV, we give a measure of rotation
of plane of polarization. In Sec.V, we show how one identifies the regions
of interest by suitably choosing the control field
parameters. In Sec. VI, we present numerical results that substantiates
the analytical results. We show that indeed large MOR could be obtained
due to the control field. We analyze different
probe frequency regions to understand the contributions of electric and
magnetic field to the large polarization rotation. In Sec. VI, we discuss
a special case where the counter propagating probe and control field are
in two-photon resonance with the $|e\rangle \leftrightarrow |g\rangle$
transition (see Fig.\ 2). We show both analytically and numerically that
this configuration can be advantageous for enhancement of MOR. We conclude
with a summary of the results in Sec.VII.

\section{The Model and The Susceptibilities}

\begin{figure}
\hspace*{-1.2 cm}
\epsfxsize 4.5in
\centerline{
\epsfbox{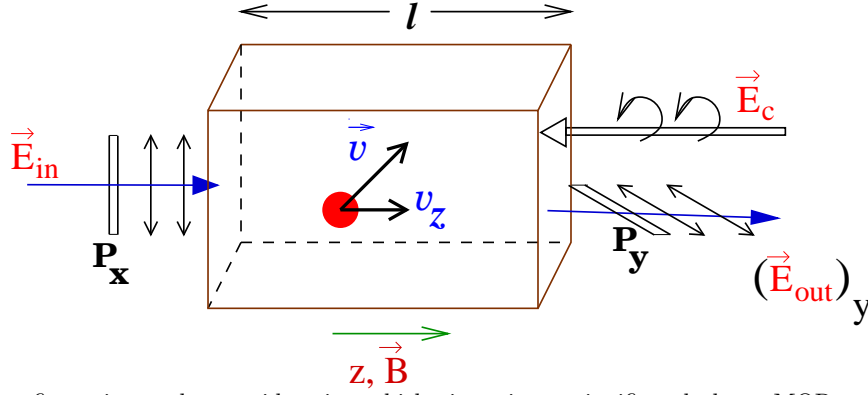}}
\caption{
The configuration under consideration which gives rise to significantly
large MOR and large enhancements. The direction of magnetic field
$\vec{B}$ fixes quantization axis ($z$-axis).
The control field ($\vec{E}_c$) and the input probe field ($\vec{E}_{in}$)
are counter propagating along the $z$-axis.
The atom in the cell moves with velocity $\vec{v}$ in arbitrary directions.
$P_x$ and $P_y$ are $x$-polarizer at input and $y$-polarized analyzer at the
output respectively. $(\vec{E}_{out})_y$ is the output probe after passing
through $P_y$.
}
\label{fig1}
\end{figure}

\begin{figure}
\epsfxsize 9cm
\centerline{
\epsfbox{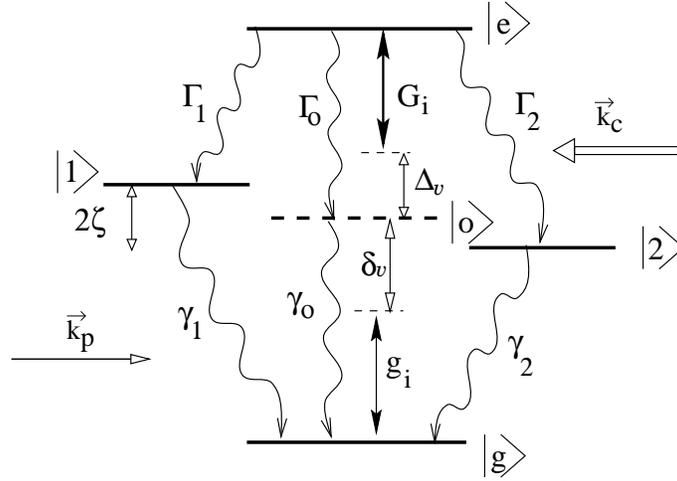}}
\caption{
The four-level model scheme having $m$-degenerate sub-levels $|1\rangle$ and
$|2\rangle$ as its intermediate states. The magnetic field $\vec{B}$ gives
rise to Zeeman splitting $2\zeta$. The spontaneous decay rates are
denoted by $2\Gamma_i$ and $2\gamma_i$. The probe field ($\vec{k}_p$) and
the control field ($\vec{k}_c$) are counter propagating. The Rabi frequencies
of the probe field and the control field are given by $2g_i$ and $2G_i$,
corresponding to the
$|i\rangle \leftrightarrow |g\rangle$ and $|e\rangle \leftrightarrow
|i\rangle$ couplings respectively ($i = 1,2$). The detunings of the probe and
the control fields from the degenerate $j=1$ state, in the moving atomic
frame of reference, are $\delta_v$ and $\Delta_v$ respectively.
}
\label{fig2}
\end{figure}

The MOR consists of the propagation of linearly polarized light
$\vec{E}_p$ tuned close to the transition $j\leftrightarrow j^\prime$
in presence of a magnetic field $\vec{B}$. The susceptibilities $\chi_\pm$
for the two circularly polarized components of the probe beam would be
different as $\vec{B} \ne 0$. We can now consider coherent control of
MOR in a configuration as depicted in Fig. \ref{fig1} with a control
field $\vec{E}_c$ which can be tuned close to another transition say
$j^\prime \leftrightarrow j^{\prime\prime}$. The atoms move randomly
inside the cell with velocity $\vec{v}$. The probe field $\vec{E}_p$ and
control field $\vec{E}_c$ are taken to be counter propagating. The model
scheme we consider (Fig. (\ref{fig2}))
is a generalization of the scheme in \cite{my_MOR}. Here we have included the
spontaneous decays between $m=0 \leftrightarrow m=0$ states in the calculation,
which was neglected in \cite{my_MOR}. The decay coefficients
corresponding to $|e\rangle \rightarrow |i\rangle$ ($|i\rangle \rightarrow
|g\rangle$) transitions are denoted by $2\Gamma_i$ ($2\gamma_i$). In what
follows below, we outline the calculation of susceptibilities of the atoms,
moving at $\vec{v}$, to the $\sigma_\pm$ components of the probe field.

    We write the fields in the circular basis as
\begin{equation}
\vec{E}_\alpha = \left( {\cal E}_{\alpha +} \hat{\epsilon}_+
+ {\cal E}_{\alpha -} \hat{\epsilon}_- \right)
e^{i\vec{k}_\alpha \cdot \vec{r} - i\omega_\alpha t}, ~~\alpha = c,~p;
\end{equation}
where
\begin{equation}
\hat{\epsilon}_\pm = \frac{\hat{x} \pm i\hat{y}}{\sqrt{2}}.
\end{equation}
Here ${\cal E}_{c\pm}$ (${\cal E}_{p\pm}$) represent the $\sigma_\mp$
components of the control field (probe field).
Let the dipole matrix elements corresponding to $|e\rangle \leftrightarrow
|i\rangle$ and $|i\rangle \leftrightarrow |g\rangle$ transitions
be represented by $\vec{D}_{ei}$ and $\vec{d}_{ig}$ respectively. The
polarization state of the incident fields decide the various field couplings
between the $j=0 \leftrightarrow j=1 \leftrightarrow j=0$ states. The
dipole matrix elements $\vec{D}_{ij}$ and $\vec{d}_{ij}$ can be written
with their corresponding Clebsch-Gordan coefficients as
\begin{eqnarray}
\vec{D}_{e1} &=& -D\hat{\epsilon}_+,~ \vec{D}_{e2} = D\hat{\epsilon}_-,
\nonumber \\
\vec{d}_{1g} &=& -d\hat{\epsilon}_-,~ \vec{d}_{2g} = d\hat{\epsilon}_+;
\label{dipoleM}
\end{eqnarray}
where $D$ ($d$) denotes the reduced dipole matrix element corresponding to
upper (lower) $j=0\leftrightarrow j=1$ ($j=1\leftrightarrow j=0$) transitions.

    In the rotating wave approximation, the interaction Hamiltonian
${\cal H}_I$ corresponding to the scheme in Fig. \ref{fig2} is
\begin{eqnarray}
{\cal H}_I (t) = -\hbar \sum_{i=1,2} \left[
|i\rangle\langle g| g_i e^{-i\omega_p t +i\vec{k}_p\cdot\vec{v} t} +
|e\rangle\langle i| G_i e^{-i\omega_c t +i\vec{k}_c\cdot\vec{v} t}
+ H.c.
\right];
\label{intH}
\end{eqnarray}
where Rabi frequencies $2G_i$ and $2g_i$ of the control and probe lasers are
\begin{equation}
G_i = \frac{\vec{D}_{ei}\cdot \vec{\cal E}_c}{\hbar},~
g_i = \frac{\vec{d}_{ig}\cdot \vec{\cal E}_p}{\hbar}.
\label{Rabif}
\end{equation}
On combining Eq.(\ref{dipoleM}) and (\ref{Rabif}) we obtain,
\begin{eqnarray}
G_1 = -\frac{D{\cal E}_{c-}}{\hbar}&,&
G_2 = \frac{D{\cal E}_{c+}}{\hbar},
\\ \nonumber
g_1 = -\frac{d{\cal E}_{p+}}{\hbar}&,&
g_2 = \frac{d{\cal E}_{p-}}{\hbar}.
\end{eqnarray}
In terms of Fig. \ref{fig2}, the unperturbed Hamiltonian ${\cal H}_0$ is
\begin{equation}
{\cal H}_0 = \hbar(\omega_{eo} +\omega_{og}) |e\rangle\langle e| +
\hbar(\omega_{og} +\zeta) |1\rangle\langle 1| +
\hbar\omega_{og} |o\rangle\langle o| +
\hbar(\omega_{og} -\zeta) |2\rangle\langle 2|.
\end{equation}
Here $\hbar\omega_{eo}$ ($\hbar\omega_{og}$) is the energy separation between
$|e\rangle$ ($|g\rangle$) and $|o\rangle$, and $2\zeta = \mu_B B/\hbar$ is
the Zeeman splitting of the degenerate levels, caused by the magnetic field
$B$. The atomic dynamics is described by the master equation
\begin{equation}
\dot{\rho} = \frac{-i}{\hbar}[{\cal H}_0 +{\cal H}_I(t) ,\rho]
- \sum_{i=o,1,2} \left( \Gamma_i \{ |e\rangle\langle e|, \rho\}
+ \gamma_i \{ |i\rangle\langle i|, \rho\}
- 2\Gamma_i \rho_{ee} |i\rangle\langle i|
- 2\gamma_i \rho_{ii} |g\rangle\langle g|\right).
\end{equation}
The second term under the summation sign represents the natural decays of
the system. The curly bracket represents the anti-commutator. The explicit
time dependence can be eliminated by making a transformation $\rho
\rightarrow \tilde{\rho}$ such that
\begin{eqnarray}
\tilde{\rho}_{ii} &=& \rho_{ii},~
\tilde{\rho}_{ig} = \rho_{ig} e^{i\omega_p t - i\vec{k}_p\cdot\vec{v} t},
\nonumber \\
\tilde{\rho}_{ei} &=& \rho_{ei} e^{i\omega_c t - i\vec{k}_c\cdot\vec{v} t},~
\tilde{\rho}_{eg} = \rho_{eg} e^{i(\omega_p +\omega_c) t -
i(\vec{k}_p +\vec{k}_c)\cdot\vec{v} t}.
\end{eqnarray}
The matrix equation for $\tilde{\rho}$ is found to be
\begin{equation}
\dot{\tilde \rho} = \frac{-i}{\hbar}[{\cal H}_{\rm eff} ,\tilde{\rho}]
- \sum_{i=o,1,2} \left( \Gamma_i \{ |e\rangle\langle e|, \tilde{\rho}\}
+ \gamma_i \{ |i\rangle\langle i|, \tilde{\rho}\}
- 2\Gamma_i \tilde{\rho}_{ee} |i\rangle\langle i|
- 2\gamma_i \tilde{\rho}_{ii} |g\rangle\langle g|\right),
\label{den-eq}
\end{equation}
with the effective Hamiltonian in the transformed frame
\begin{eqnarray}
{\cal H}_{\rm eff} &=& \hbar (\delta_v + \Delta_v) |e\rangle\langle e|
+ \hbar (\delta_v + \zeta)
|1\rangle\langle 1| + \hbar (\delta_v - \zeta)|2\rangle\langle 2|
\nonumber \\
&-& \hbar
\sum_{i=1,2} \left( g_i|i\rangle\langle g| + G_i |e\rangle\langle i| + H.c.
\right);
\label{hamiltonian}
\end{eqnarray}
where
\begin{equation}
\delta_v = \delta + k_p v_z,~ \Delta_v = \Delta - k_c v_z.
\end{equation}
Here $\delta = \omega_{og} -\omega_p,~ \Delta = \omega_{eo} - \omega_c$
correspond to the detunings of the probe and control field when the atom is
stationary.  Further we assume $k_p \approx k_c$ for simplicity. Thus
one can write
\begin{equation}
\Delta_v +\delta_v \approx \Delta +\delta.
\label{dop-cancel}
\end{equation}
Here it may be noted that due to our particular choice of counter propagating
probe and control fields, the two-photon resonant terms can be made
independent of atomic velocity [See e.g. in Eq.\ (\ref{chiredp})].
The configuration consisting of counter propagating probe and control field
in ladder system has been shown to be useful in Doppler free polarization
spectroscopy \cite{hansch:76}, EIT \cite{eit} and LWI \cite{lwi}.

Let $\chi_+$ ($\chi_-$) be the susceptibilities of the moving atom
corresponding to the $\sigma_-$ ($\sigma_+$) component of the probe field.
We choose the probe field polarization such that $g_i \ne 0$.
One can write $\chi_\pm$ in terms of dimensionless quantities as
\begin{equation}
\chi_\pm = \left( \frac{\alpha}{4\pi k_p}\right) s^\pm;
\label{chi}
\end{equation}
where $s^\pm$, the normalized susceptibilities are given by
\begin{equation}
s^+ = \left(\frac{\tilde{\rho}_{1g}\gamma}{g_1}\right),
s^- = \left(\frac{\tilde{\rho}_{2g}\gamma}{g_2}\right) .
\end{equation}
Here $\alpha l$ is weak probe field absorption at the line center and is given
by $\alpha l = 4\pi k_p l |d|^2 n/(\hbar \gamma)$; where $n$ denotes the
atomic density and $l$ is the length of the cell.
For simplicity, we assume $\gamma_1 = \gamma_2 = \gamma_o =\gamma$. Under
steady state
conditions, we solve Eq.\ (\ref{den-eq}) to obtain complete analytical
solutions for the susceptibilities $\chi_\pm$ or the normalized
susceptibilities $s^\pm$
\begin{equation}
s^+ = \frac{i\gamma \left[ |G_2|^2 +{\bf (}\gamma +i(\delta_v
-\zeta){\bf )(}\Gamma_o +\Gamma_1+\Gamma_2 +i(\Delta_v+\delta_v){\bf )}\right]}
{|G_2|^2 {\bf (}\gamma +i(\delta _v+\zeta){\bf )}
+{\bf (}\gamma+i(\delta _v-\zeta){\bf )}
\left[ |G_1|^2 +{\bf (}\gamma +i(\delta _v+\zeta){\bf )}
{\bf (}\Gamma_o+\Gamma_1 +\Gamma_2 +i(\Delta_v+\delta_v){\bf )}
\right]},
\label{res1}
\end{equation}
\begin{equation}
s^- = \frac{i\gamma \left[ |G_1|^2 +{\bf (}\gamma +i(\delta_v
+\zeta){\bf )(}\Gamma_o+\Gamma_1 +\Gamma_2 +i(\Delta_v +\delta_v){\bf )}\right]}
{|G_1|^2 {\bf (}\gamma +i(\delta_v -\zeta){\bf )}
+ {\bf (}\gamma +i(\delta_v +\zeta){\bf )}
\left[ |G_2|^2 +{\bf (}\gamma +i(\delta_v -\zeta){\bf )}
{\bf (}\Gamma_o+\Gamma_1 +\Gamma_2 +i(\Delta_v +\delta_v){\bf )}
\right]}.
\label{res2}
\end{equation}
In writing (\ref{res1}) and (\ref{res2}), we have used the condition
(\ref{dop-cancel}). We note that the atomic velocity dependence of $s^\pm$
comes via $\delta_v$. The results presented above are susceptibilities
of the atoms moving at $\vec{v}$, to the lowest order in the probe field.
The response of the medium to the input probe field can be obtained by
averaging $s^\pm$ over the distribution of velocities.
It may be noted that the parameter space in Eqs.\ (\ref{res1}) and (\ref{res2})
is very large. Therefore we {\em identify} a particular configuration of our interest
and work only in the region which gives
large asymmetry between $\langle s^+ \rangle$ and $\langle
s^- \rangle$ ($\langle~ \rangle$ represents average over the velocity 
distribution of atoms inside the cell), and can lead to large MOR. We
focus on a particularly interesting case when $G_2 = 0$;
i.e., the control field is $\sigma_+$-polarized
(${\cal E}_c \equiv {\cal E}_{c-} \ne 0$ and ${\cal E}_{c+} =0$) and it 
couples to the $|1\rangle \leftrightarrow |e\rangle$ transition only. Clearly 
$s^-$ becomes
\begin{equation}
s^- = \frac{i\gamma}{{\bf (}\gamma +i(\delta_v -\zeta){\bf )}};
\label{chiredm}
\end{equation}
which is independent of the control field parameters.
Whereas $s^+$ is strongly dependent on the strength and
frequency of the control field and is given by
\begin{equation}
s^+ = \frac{i\gamma{\bf (}\Gamma_o+\Gamma_1 +\Gamma_2 + i(\Delta + \delta)
{\bf )}}
{|G_1|^2 + {\bf (}\gamma +i(\delta_v+\zeta){\bf )(}\Gamma_o+\Gamma_1
+\Gamma_2 + i(\Delta + \delta){\bf )}}.
\label{chiredp}
\end{equation}
In absence of the control field, the susceptibilities reduce to
\begin{equation}
s^\pm = \frac{\gamma}{{\bf (}(\delta_v \pm\zeta) -i\gamma{\bf )}};
\label{chi-no-fld_dop}
\end{equation}
which clearly indicates that $s^\pm$ are completely symmetric in absence
of magnetic field (i.e. $\zeta = 0$). Most of the MOR studies with a weak
coherent field use the susceptibility in (\ref{chi-no-fld_dop}). Further
it may be noted that from Eq. (\ref{chiredm}) and Eq. (\ref{chiredp}) that,
$s^+ \ne s^-$ even in absence of magnetic field when $G_1 \ne 0$. This
explains the laser induced birefringence reported by Winelandy and Gaeta
\cite{gaeta:98}.

\section{Susceptibilities $\chi_\pm$ of the Doppler Broadened Medium}

Next we calculate the $\chi_\pm$ of a Doppler broadened medium. Here,
as mentioned in Sec. II, one needs to average $s^\pm$
over the atomic velocity distribution $\sigma (v_z)$ inside the cell
to obtain the response of the medium
\begin{equation}
\langle s^\pm\rangle = \int_{-\infty}^{\infty} s^\pm (v_z)
\sigma (v_z) dv_z.
\label{spm_av_int}
\end{equation}
It is assumed that at thermal equilibrium, the atoms inside the cell follow
Maxwell-Boltzmann velocity distribution
\begin{equation}
\sigma (v_z) = (2\pi K_B T/M)^{-1/2} \exp(-Mv_z^2/2K_B T),
\end{equation}
where mass of the moving atom is $M$, temperature of the cell $T$ and
$K_B$ is Boltzmann constant.
For convenience, transforming the integral in Eq.\ (\ref{spm_av_int}) from velocity
space to frequency space \cite{demtroder}, we get
\begin{equation}
\langle s^\pm \rangle
= \int_{-\infty}^{\infty} s^\pm (\delta_v)
\sigma(\delta_v) d\delta_v~,
\label{dop-av}
\end{equation}
where the distribution in frequency space is
\begin{equation}
\sigma(\delta_v) \equiv \frac{1}{\sqrt{2\pi\omega_D^2}}
\exp \left[ -(\delta_v -\delta)^2 / 2\omega_D^2 \right]; ~
\omega_D \simeq \omega_{og} (K_B T/Mc^2)^{\frac{1}{2}}.
\label{maxwell-freq}
\end{equation}
Here $\omega_D$ represents the Doppler width in frequency space. For our
case of $\sigma_+$ polarized control field, we
substitute $s^\pm$ from Eqs.\ (\ref{chiredm}) and (\ref{chiredp})
in Eq.\ (\ref{dop-av}) and
evaluate the integral. We could obtain the complete analytical results
for the Doppler averaged susceptibilities, in terms of complex
error functions \cite{abbr_stegun} as
\begin{eqnarray}
\langle s^-\rangle &\equiv& \frac{i\pi\gamma}{\sqrt{2\pi \omega_D^2}}
{\cal W}\left( \frac{\zeta -\delta +i\gamma}{\sqrt{2}\omega_D}\right);~
\label{chiavm}
\\
\langle s^+\rangle &\equiv& \frac{i\pi\gamma}{\sqrt{2\pi \omega_D^2}}
{\cal W}(\xi);
\label{chiavp}
\end{eqnarray}
\begin{equation}
\xi = \frac{1}{\sqrt{2}\omega_D}\left[i\gamma -\zeta -\delta
+ \frac{|G_1|^2}{\Delta +\delta -i(\Gamma_o+\Gamma_1 +\Gamma_2)}\right].
\end{equation}
The ${\cal W}$ function is defined as
\begin{equation}
{\cal W}(z) = \frac{i}{\pi}\int_{-\infty}^{\infty}
\frac{e^{-t^2} dt}{z -t}.
\end{equation}
It can be written in terms of the error function ${\rm Erf}(z)$ as
\begin{equation}
{\cal W}(\alpha) = e^{-\alpha^2} (1 - {\rm Erf}(-i\alpha));~
{\rm Erf} (z) = \frac{2}{\sqrt{\pi}}\int_0^z e^{-t^2} dt.
\end{equation}
It may be noted that the argument of ${\cal W}$ function in
$\langle s^- \rangle$ will show
usual Doppler profile since it is independent of the control field
but the argument of ${\cal W}$ function in $\langle s^+ \rangle$ is strongly dependent on the strength and
frequency of the control field and therefore, the Doppler profile can
be modified with these control field parameters.

\section{Measure of Rotation}

        Using the $\langle s^\pm\rangle$ obtained above, the rotation of
polarization $\theta$ of the probe can be determined from Eq.(1) which, however,
is valid only if the absorption of the medium is very small. Since we
consider the resonant or near-resonant MOR, one also needs to take into
account the large absorption associated with the large dispersions near
resonance. Absorption contributes to the polarization rotation via dichroism
(rotation solely due to Im $\langle s^\pm\rangle$)
but large absorption attenuates the MOR signal at the output.

    Let us consider an $x$-polarized incident probe field propagating
along the quantization axis $z$. The field amplitude can be written as
\begin{equation}
\vec{\cal E}_{in} = \vec{\cal E}_{p} (z=0) = \hat{x} {\cal E}_0;
\end{equation}
which can be resolved into two circularly polarized components as
\begin{eqnarray}
\vec{\cal E}_{in} &=&
\hat{\epsilon}_+ {\cal E}_{p+} (z=0) + \hat{\epsilon}_-
{\cal E}_{p-} (z=0)
\nonumber \\
&=& \frac{{\cal E}_0}{\sqrt{2}} (\hat{\epsilon}_+ + \hat{\epsilon}_-).
\end{eqnarray}
When the probe field $\vec{\cal E}_{in}$ passes through the anisotropic medium,
${\cal E}_{p\pm} (z)$ evolves. In the limit of a
weak probe, we get the output field
\begin{equation}
\vec{\cal E}_{out} = \vec{\cal E}_p (z=l) =
\frac{{\cal E}_0}{\sqrt{2}}
\left[ \hat{\epsilon}_+
e^{\left( i\frac{\alpha l}{2} \langle s^+\rangle\right)}
+ \hat{\epsilon}_-
e^{\left( i\frac{\alpha l}{2} \langle s^-\rangle \right)}\right].
\label{output}
\end{equation}
Clearly, $\vec{\cal E}_{out}$ contains both $x$ and $y$-polarization
components, and thus polarization of the probe is rotated. For small
absorption, it is easy to derive the rotation angle $\theta$ in Eq.(1).
Experimentally one observes the rotation by measuring the intensity after
passing the output through a crossed polarizer $P_y$ (as shown in Fig.1)
given by
\begin{equation}
T_y = \frac{|({\cal E}_{out})_y|^2}{|{\cal E}_{in}|^2}
= \frac{1}{4} \left| \exp \left( i\frac{\alpha l}{2} \langle s^+\rangle\right)
- \exp \left( i\frac{\alpha l}{2} \langle s^-\rangle\right) \right|^2;
\label{ty}
\end{equation}
which gives the measure of polarization rotation of the weak $x$-polarized
probe field. Here the intensity of transmission through $P_y$, is scaled
with the input intensity in $x$-polarization. It should be borne in mind
that $\langle s^\pm \rangle$ are in general complex.

\section{Condition for Enhancement of MOR}

In this section we {\em identify} the regions of our interest. We determine
the criteria to choose the control field parameters to efficiently control and
hence enhance the MOR. From Eq.\ (\ref{ty}), one observes the following:

\noindent
(i) When $\langle s^+ \rangle \approx \langle s^- \rangle$,
$T_y \rightarrow 0$.

\noindent
(ii) When  Re $\langle s^+ \rangle \simeq$ Re $\langle s^- \rangle$ but
Im $\langle s^+ \rangle \ne$  Im $\langle s^- \rangle$, $T_y$ reduces to
\begin{equation}
T_y \simeq \frac{1}{4} \left| e^{\left(-\frac{\alpha l}{2} {\rm Im}\langle s^+ \rangle
\right)}
- e^{\left(-\frac{\alpha l}{2} {\rm Im} \langle s^- \rangle \right)}\right|^2.
\end{equation}
If both $\frac{\alpha l}{2} {\rm Im} \langle s^\pm \rangle$ are large,
$T_y \rightarrow 0$. However if $\frac{\alpha l}{2} {\rm Im}\langle s^+ \rangle$
is large but $\frac{\alpha l}{2} {\rm Im} \langle s^- \rangle$ is small
(or vice versa), we obtain
\begin{equation}
T_y \simeq \frac{e^{\alpha l~{\rm Im}\langle s^- \rangle}}{4}~\rightarrow \frac{1}{4};
\end{equation}
which is the rotation due to {\it dichroism only}.

\noindent (iii) Further {\em when Im $\langle s^+ \rangle \approx$
Im $\langle s^- \rangle = \beta$ (say) but Re $\langle s^+ \rangle
\ne$ Re $\langle s^- \rangle$}, we get
\begin{equation}
T_y \simeq \frac{e^{(-\alpha l\beta)}}{4} \left| 1 -
e^{i\frac{\alpha l}{2} {\rm Re} (\langle s^- \rangle -\langle s^+
\rangle)} \right|^2.
\end{equation}
If $\alpha l\beta$ is small,
\begin{equation}
T_y \simeq \frac{1}{4} \left| 1 - e^{i\frac{\alpha l}{2} {\rm Re}(\langle s^-\rangle
- \langle s^+\rangle)} \right|^2,
\label{ty_condn1}
\end{equation}
thus when
\begin{equation}
\frac{\alpha l}{2} {\rm Re} (\langle s^-\rangle
- \langle s^+\rangle) = (2n+1)\pi~~ (n=0,1,2,...), ~ T_y = 1.
\label{condn}
\end{equation}
This is the {\em most useful region} for our system. This rotation is solely
due to birefringence. However if $\alpha l \beta$ is large then
$T_y \rightarrow 0$. This is because a large attenuation of the MOR signal
occurs. Thus we have identified that {\em the most
interesting frequency region corresponds to very small value of
${\rm Im}\langle s^\pm\rangle$ and when the asymmetry between
${\rm Re}\langle s^\pm\rangle$ satisfies the condition (\ref{condn})}.
Therefore our objective is to select proper
control field parameters so that above condition can be achieved.

\section{Numerical Results on Coherent Control of MOR}

    Based on the above observations, we present some interesting
numerical results for different parameters to demonstrate the large
enhancement of MOR. We define the MOR signal enhancement factor
\begin{equation}
\eta = \frac{\left( T_y \right)_{G_1\ne 0}}{\left( T_y \right)_{G_1 = 0}}.
\end{equation}
For a given $\delta$, $\eta$ represents the enhancement (if $\eta > 1$) or
suppression (if $\eta < 1$)
of MOR signal by a control field, when compared to the MOR without control
field. We use the notation $\langle s^\pm_0 \rangle$ to represent
susceptibilities corresponding to $\sigma_\mp$ components of the probe when
control field is absent and $\langle s^+_c \rangle$ to represent the
susceptibility modified by the control field. In the following we give some
typical values of various physical parameters used here:
the Doppler width $\omega_D = 50\gamma$ corresponds to $^{40}Ca$ cell at a
temperature of $\sim 500K$. For length of the cell $l = 5cm$, $\alpha l =300$
corresponds to an atomic density of $\sim 10^{12}~atoms/cm^3$, a Zeeman
splitting of $2\zeta =20\gamma$ corresponds to a magnetic field of strength
$\sim 200~Gauss$, and $G_1 = 100\gamma$ would correspond to a laser field
of strength $\sim 5~W/cm^2$.

\begin{figure}
\epsfxsize 7cm
\centerline{
\epsfbox{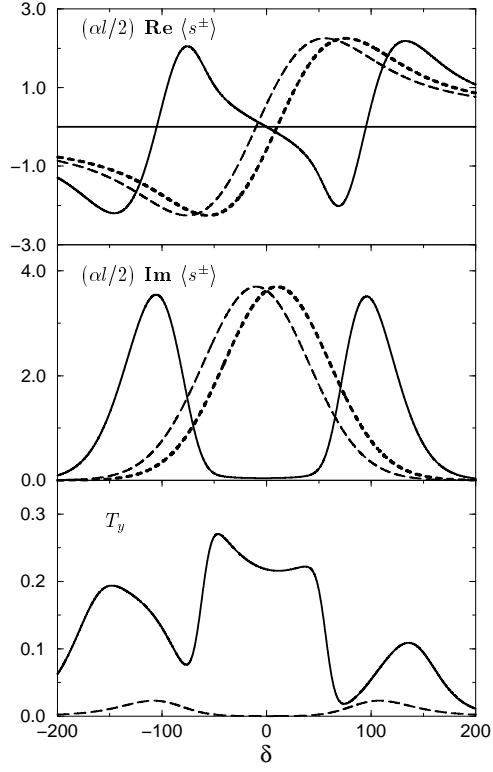}}
\caption{
The enhancements of MOR for a control field tuned to $|e\rangle
\leftrightarrow |o\rangle$ transition ($\Delta = 0$) with Rabi frequency
$G_1 = 100$. In the plots for $\frac{\alpha l}{2}\langle s^\pm\rangle$,
the thick-dashed (long-dashed) lines represent $\frac{\alpha l}{2}\langle
s^-_0 \rangle$ ($\frac{\alpha l}{2}\langle s^+_0 \rangle$)
and solid lines represent $\frac{\alpha l}{2}\langle s^+_c \rangle$.
In the plot
for $T_y$, dashed (solid) curve represents the rotation without (with)
control field. The other parameters used are $\omega_D = 50,~\zeta = 10$
and $\alpha l = 300$. All frequencies are scaled with
$\Gamma_o = \Gamma_1 = \Gamma_2 = \gamma$.
}
\label{fig3}
\end{figure}

        In Fig. \ref{fig3}, we consider the effect of the control field
which is on resonance with the transition
$|e\rangle\leftrightarrow|o\rangle$ (i.e. $\Delta = 0$). We
consider density of atoms in the cell such that $\alpha l = 300$.
We observe significant enhancement of MOR for a large range of
probe frequencies. (i) We get the enhancement factor $\eta =
1.04\times 10^3$ for $\delta = 0$. This can be understood as
follows: in the absence of the control field and for $\delta = 0$,
${\rm Im}\langle s^+_0\rangle = {\rm Im}\langle s^-_0\rangle =
\beta$ (say) and $\alpha l \beta$ is large, leading to $T_y
\approx 0$ due to large signal attenuation by absorption [see Eq.\
(\ref{ty_condn1})]. By application of a control field, the
absorption peak (Im $\langle s^+_0 \rangle$) splits - giving rise to
Autler-Townes doublet. The minimum of ${\rm Im} \langle
s^+_c\rangle$ appears at $\delta \sim 0$. Thus MOR signal at this
frequency is enhanced by suppressing the $\sigma_-$ component of
the probe field as a result of its large absorption. (ii) Further,
large MOR signal is observed for a fairly large range of probe
frequencies ($-50 < \delta < 50$) - which is attributed to the
flipping of the sign of ${\rm Re} \langle s^+\rangle$ causing a
larger asymmetry between Re$\langle s^+_c\rangle$ and Re$\langle
s^-_0\rangle$. However large absorption reduces most part of the
rotation signal. Hence it is observed that $T_y$ is maximum
($\approx 27\%$ at $\delta \approx -50$) when both Im $\langle s^-
\rangle$ and Im $\langle s^+_c \rangle$ are small.

\begin{figure}
\epsfxsize 3.5in
\centerline{
\epsfbox{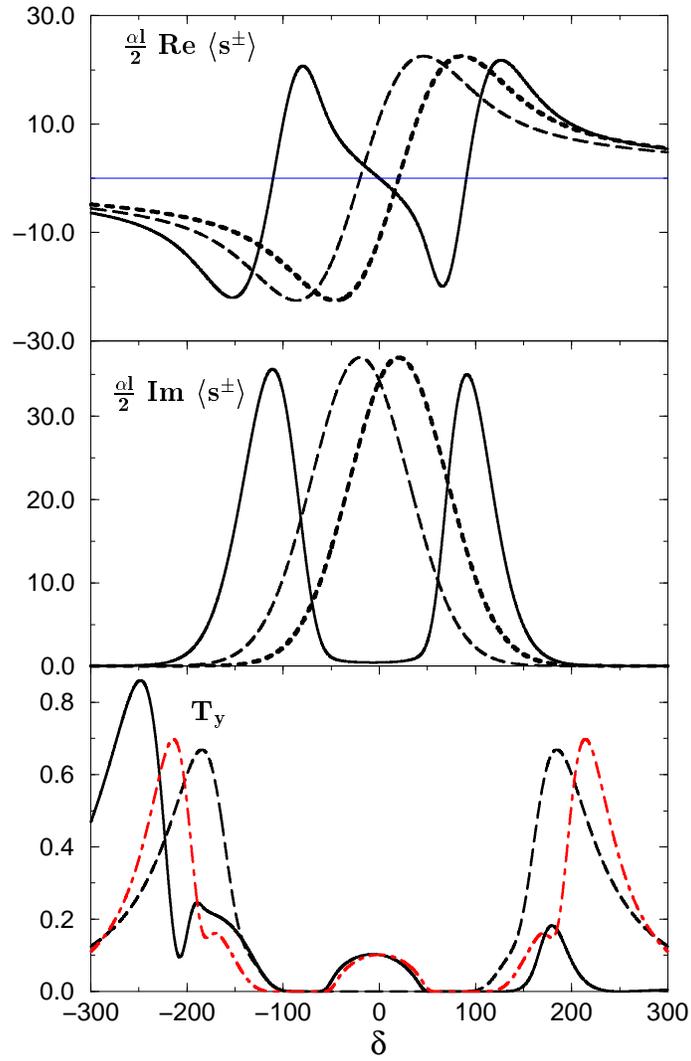}}
\caption{
Enhancement and suppression of MOR in a denser medium with $\alpha l = 3000$.
The legends of the curves used are same as in Fig. \ref{fig3}. The magnetic field used
in this plot is $\zeta = 20$, the control field Rabi frequency is taken to be
$G_1 = 100$ and the Doppler width is $\omega_D = 50$. A plot of $T_y$ with
$\zeta = 0$ but with ${\cal E}_c \ne 0$ (dot-dashed line in the plot for $T_y$)
is also presented to isolate the roles
of ${\cal E}_c$ and $B$ in controlling the MOR. All frequencies are scaled with
$\Gamma_o = \Gamma_1 = \Gamma_2 = \gamma$.
}
\label{fig4}
\end{figure}

In Fig. \ref{fig4}, we consider the control of MOR in a denser medium. Here
$\alpha l = 3000$ and the magnetic field is such that $\zeta = 20$.
In order to demonstrate the combined effect of ${\cal E}_c$ and $B$, and then
to isolate the contribution of magnetic field in obtaining large $T_y$,
we have also plotted $T_y$ for $B = 0$ but ${\cal E}_c \ne 0$.
In the following we discuss the contribution of ${\cal E}_c$ and $B$ in
different probe frequency regions. To understand the enhancements and
suppressions of the MOR signal at different probe frequencies,
we analyze the following different regions in Fig.\ \ref{fig4} :

\noindent
{\em Region I:} For $-50 < \delta < 50$, Im $\langle s^\pm_0 \rangle$ are
large. Thus in the absence of control field, $T_y$ in this region is almost
zero. However by application of control field, an absorption minimum for
$\sigma_-$ polarization component ({\rm Im} $\langle s^+_c\rangle$) occurs due to
EIT at $\delta = 0$. Thus a large enhancement of MOR is obtained when
${\cal E}_c \ne 0$
compared to the case of ${\cal E}_c = 0$. However $T_y$ value is only
$10.2\%$ of the input probe intensity at $\delta = 0$, because
$\frac{\alpha l}{2}{\rm Im} \langle s^-_0 \rangle$ still has a large value and
therefore
\begin{equation}
T_y \approx \frac{1}{4}
\left| e^{i\frac{\alpha l}{2} \langle s^+_c \rangle} \right|^2
\approx \frac{1}{4} e^{-\alpha l {\rm Im} \langle s^+_c \rangle}
\end{equation}
is small. This rotation is solely due to {\em dichroism}
created by the control laser. Comparing the $T_y$ values with
$B = 0, {\cal E}_c \ne 0$ (dot-dashed line) and $B \ne 0, {\cal E}_c \ne 0$
(dashed line), it is clear from the Fig. \ref{fig4} that the magnetic
field contribution is very small in this region.

\noindent
{\em Region II:} In the region $-100 < \delta < -50$,
there are residual absorptions at the tail of the Lorentzian
Im $\langle s^\pm_0 \rangle$. Further Im $\langle s^+_c \rangle$ is also
large in this region. Therefore though
there is a large asymmetry between Re $\langle s^-_0 \rangle$ and
Re $\langle s^+_c \rangle$, very large attenuation makes the value of
$T_y$ extremely small.

\noindent
{\em Region III:} In the probe frequency region $-200<\delta <-100$,
minimum of Im $\langle s^\pm_0 \rangle$ occurs but
Im $\langle s^+_c \rangle$ still has large value in this region. Thus the
rotation is large in absence of ${\cal E}_c$ but with the control field,
there occurs a large suppression of the MOR signal.

\noindent {\em Region IV:} For $-300<\delta <-200$, we get the
most interesting region because the Im  $\langle s^\pm_0 \rangle$
and Im $\langle s^+_c \rangle$ are very small. Thus even though
the asymmetry between Re $\langle s^\pm \rangle$ is small, the
birefringence contribution shows up in the form of very large
rotation in this region. For example, the MOR signal at $\delta =
-248.3$ is $86.1\%$ of the input intensity. The comparison of
control field induced $T_y$ in presence and absence of magnetic
field clearly demonstrates that the presence of magnetic field
causes larger asymmetry between $\langle s^+_c \rangle$ and
$\langle s^-_0 \rangle$ in this region. For example at $\delta
\approx -300$, $T_y$ with magnetic field is about 5 times larger
compared to that without magnetic field. However in the +ve
$\delta$ region ($200<\delta <300$), the asymmetry between
$\langle s_0^-\rangle$ and $\langle s^+_c\rangle$ is reduced, and
hence MOR is suppressed.

\begin{figure}
\epsfxsize 7cm
\centerline{
\epsfbox{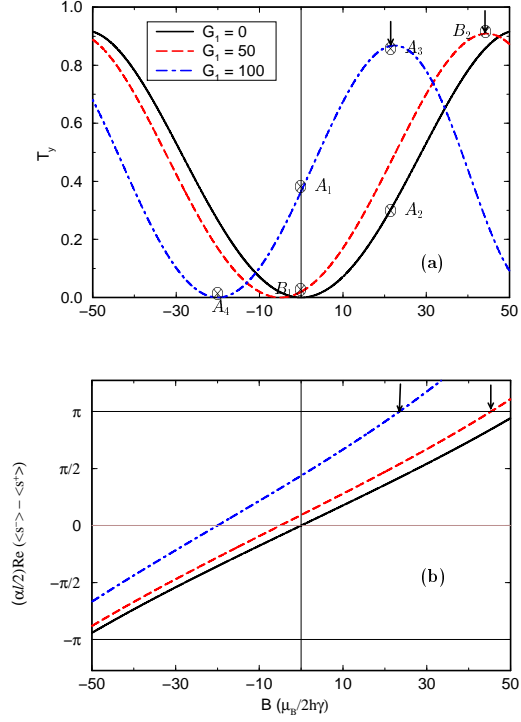}}
\caption{
(a) The plot of $T_y$ as a function of $B$ to investigate the role of magnetic
field. This plot corresponds to $\delta = -250$ (in the region IV of
Fig. \ref{fig4}).  All other parameters are same as in
Fig. \ref{fig4}. (b) The asymmetry between $\langle s^+\rangle$
and $\langle s^-\rangle$ is plotted as a function of $B$ corresponding to the
plots of $T_y$ in (a). The points marked by the arrows satisfy the condition
for maximal rotation (\ref{condn}).
\label{fig5}
}
\end{figure}

In order to bring out the role of magnetic field in the enhancement of $T_y$
observed in this region, we present Fig. \ref{fig5}(a) - where $T_y$ vs
magnetic field is plotted with a probe frequency fixed
($\delta = -250$) in the region {\em IV}. The figure clearly
demonstrates the contribution of magnetic field and laser field separately
in the enhancement of $T_y$. For clarity of the explanation, we have marked
some points in the graph. The point $A_1$ ($B_1$) represents the
rotation due to control field alone with $G_1 = 100$ ($G_1 = 50$).
The points $A_2$
($A_3$) gives the amount of $T_y$ without (with) the control field for
a given value of $\zeta = 22.4$ ($\zeta =44.45$). Thus clearly, $A_3$
represents enhancement of rotation by a factor of 2.37 due to the magnetic
field with respect to $A_1$, and when compared with $A_2$, the point
$A_3$ represents enhancement due to the control field by a factor of 2.66.
Very large $T_y (\approx 86.8\%$ of input intensity) is obtained
for $\zeta = 22.4$. The plot with $G_1 =50$ shows a large $T_y$
($\approx 90.9\%$ of input intensity) value at $\zeta = 44.54$ which
corresponds to an
enhancement of $4.5\times 10^3$ times the value compared to the point $B_1$.
Similarly large suppression of MOR can be observed when the magnetic field
is flipped (i.e. $\zeta$ is negative); e.g., the point $A_4$. The large
MOR signals and enhancements described above are interpreted by the condition
(\ref{condn}). {\em The points where the condition (\ref{condn}) is satisfied
are marked by arrows in Fig. \ref{fig5}(b)}. The Fig. \ref{fig5}(b)
also depicts the parameters for which the rotations are optimal. From
Fig. \ref{fig5}, it may be noted that large $T_y$ can be produced either
by a large magnetic field
and a weaker control field, or a weaker magnetic field and a strong control
field. This could be advantageous as
it is difficult to produce large magnetic fields in laboratory. Further using
the large enhancements ($\eta$) of MOR it is possible to realize a
{\em magneto-optical switch}, that can switch the incident polarization of the
probe to its orthogonal polarization \cite{my_DOP_springer,kerr}.

\section{MOR in Two-photon Resonance Condition}

        In this section we consider the enhancement of MOR when the
$\sigma_+$ polarized control filed and the probe field are always on two-photon
resonance with $|e\rangle \leftrightarrow |g\rangle$ transition
($\Delta +\delta = 0$). In the following discussion, we consider both the
cases of stationary atom and homogeneously broadened atom with the above
condition.

\noindent
1. {\bf Stationary Atom Case:}

        For a stationary atom, the susceptibilities are given by Eq.\
(\ref{chiredm}) and (\ref{chiredp}) but with $\delta_v \rightarrow \delta$.
Under the condition $\delta + \Delta = 0$, $s^+_c$ (that denotes $s^+$ in
presence of control field) reduces to
\begin{equation}
s^+_c = \frac{i\gamma}{\left( \frac{|G_1|^2}{\Gamma_o+\Gamma_1+\Gamma_2}
+\gamma\right) +i(\delta+\zeta)};
\label{sp_2ph}
\end{equation}
which is a Lorentzian profile with a width given by
$\left(\frac{|G_1|^2}{\Gamma_o+\Gamma_1+\Gamma_2} +\gamma\right)$.
The width is too large for a control field $G_1 \gg \gamma$,
causing large power broadening. Thus for small values of $\delta+\Omega$,
$s^+_c$ is negligibly small. However, $s^-$ remains
unchanged. Therefore $T_y$ (in Eq.(34))
reduces to
\begin{equation}
T_y = \frac{1}{4} \left| 1 - e^{i\frac{\alpha l}{2} s^-}\right|^2;
\end{equation}
and hence $T_y$ becomes independent of the control field for $|G_1| \gg
|\delta +\zeta|$. The $(T_y)_{max}$ value thus
remains same for any arbitrary value of $\zeta$; e.g. for $G_1 = 20$,
$(T_y)_{max} \sim 60\%$ for any $\zeta$ (results not shown here).
However changing the magnetic field, the $T_y$ structure shifts along
$\delta$ and $T_y$ curve is symmetric about $\zeta$.
Therefore by choosing proper magnetic field , one can
produce large $T_y$ and enhancement of $T_y$ at the required probe frequency
regions.

2. {\bf Doppler Broadened Case:}

We further consider the enhancement of MOR in the Doppler broadened
medium with the fields ${\cal E}_p$ and ${\cal E}_c$ in two-photon resonance
condition, where $\Delta + \delta = 0$. Under this condition
$\langle s^+ \rangle$ in Eq. (27) is modified which contains the control
field parameters but $\langle s^- \rangle$ remains unchanged.
Further in the limit $|G_1| \gg \omega_D$, one can show that
$\langle s^+ \rangle$ becomes equal to the corresponding stationary
atom value of $s^+$ in Eq.\ (\ref{sp_2ph}). However, note that
$\langle s^- \rangle$ [in Eq.\ (\ref{chiavm})] is still velocity dependent.
In the above limit, large power broadening is introduced in $\langle s^+
\rangle$ and amplitude of $\langle s^+_c \rangle$ is reduced. However,
this turns out to be advantageous, particularly because large asymmetry is
created between Re $\langle s^+_c \rangle$ and Re $\langle s^-\rangle$ around
$\delta = \zeta$. Moreover the absorption Im $\langle s^-_c \rangle$
becomes extremely small. Since $\langle s^-\rangle$ is Doppler broadened and
$\langle s^+_c \rangle$ is reasonably small and flat for a broad range of
$\delta$, one obtains {\em large enhancement
of $T_y$ for a broad range of probe frequencies compared to the
homogeneously broadened case}. Further, MOR in two-photon resonance condition
turns out to be advantageous for
smaller magnetic fields where very large enhancement of MOR is obtained.

\section{Summary}
In summary, we have shown how a control field can be used to control birefringence
and hence enhance MOR in a Doppler broadened medium.
We have shown how control laser can modify the susceptibilities and hence
result significantly large MOR in frequency regions, where MOR otherwise
is small. The key to large enhancement of MOR consists of utilizing the
large asymmetry in the susceptibilities caused by
the Autler-Townes splitting. We have derived conditions to select
frequency regions where one can obtain large MOR. The most useful regions
are found to be at the probe frequencies - where absorptions of both the
circularly polarized components are negligible and the associated dispersions
are quite different. We have substantiated these
analytical results using many numerical plots for many different
parameters at different conditions. We have also demonstrated the
significance of magnetic field and control field in obtaining the large MOR
by isolating the effects of the two fields. Finally we have discussed the
possibility of large enhancement of MOR for a broad range of probe frequencies
- when probe and control fields are in two-photon resonance condition.

\begin{figure}
\epsfxsize 7cm
\centerline{
\epsfbox{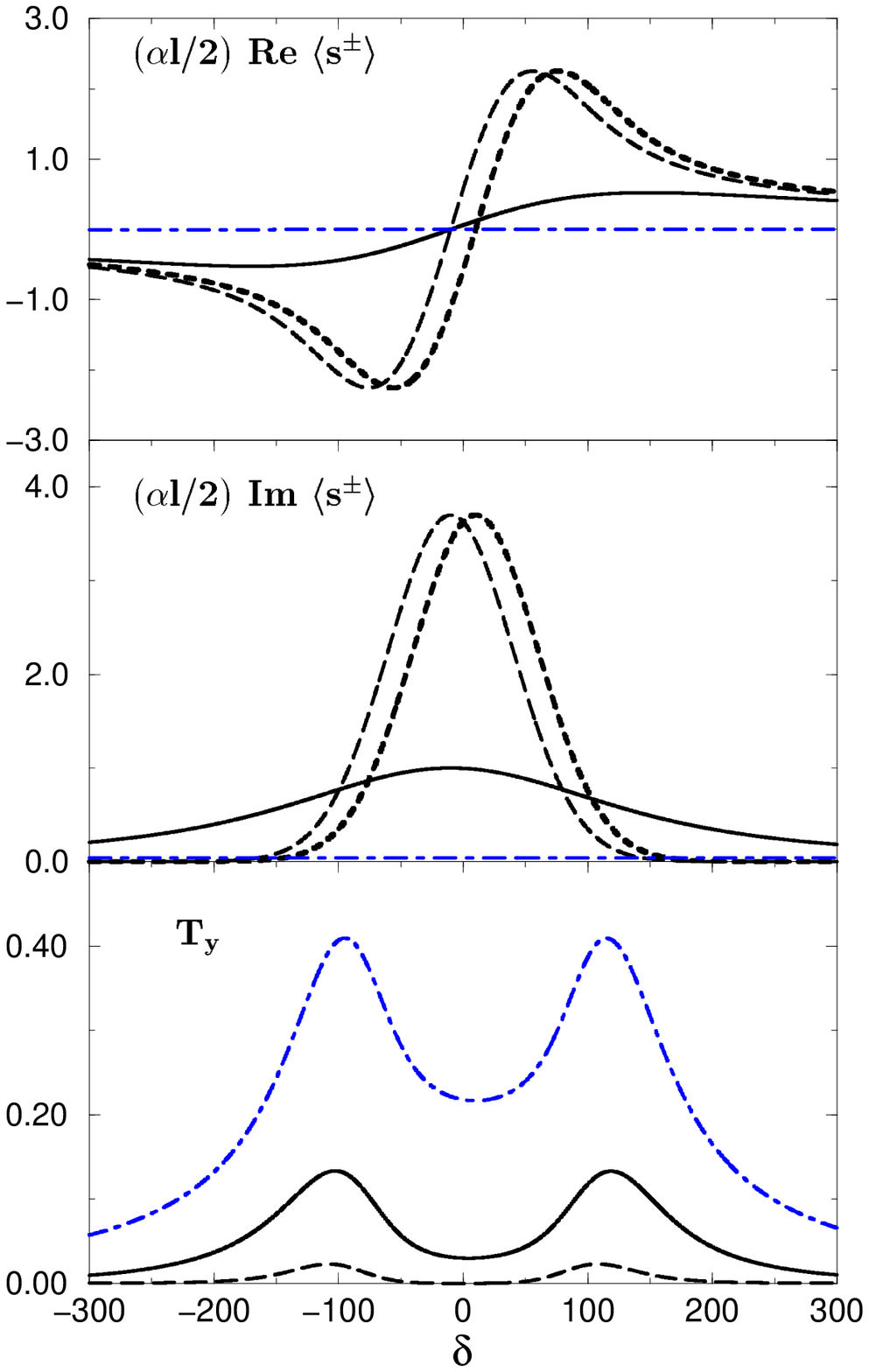}}
\caption{
Enhancement of MOR in a Doppler broadened medium when the control field
and the probe field are on two-photon resonance with $|e\rangle \leftrightarrow
|g\rangle$ transition. The legends used are same as in Fig. \ref{fig3},
the solid line corresponds to $G_1 = 20$ and the dot dashed line corresponds
to $G_1 = 100$. Here $G_1 = 100$ corresponds to the limit where $\langle
s^+_c \rangle$ becomes equal to $s^+$ in (\ref{sp_2ph}).
The other parameters used in the plot are $\zeta = 10$ and $\alpha l = 300$.
}
\end{figure}


\begin{thebibliography}{99}

\bibitem{kliger_book}
D. S. Kliger, J. W. Lewis, and C. E. Randall, ``{\em Polarized
light in optics and spectroscopy}" (Academic Press, 1990).

\bibitem{MOR_synch}
J-P Connerade, J. Phys. B {\bf 16}, 399 (1983);
J-P Connerade, T. A. Stavrakas, and M. A. Baig, {\it Synchrotron Radiation
Sources and their Applications}, ed. G. N. Greaves and I. H. Munro
(Edinburgh: SUSSP Publications, Edinburgh University, 1989), p. 310;
X. H. He and J-P Connerade, J. Phys. B {\bf 26}, L255 (1993).

\bibitem{MOR_laser}
W. Gawlik, J. Kowalski, R. Neumann, H. B. Weigemann, and K.
Winkler, J. Phys. B {\bf 12}, 3873 (1979); X. Chen, V. L. Telegdi,
and A. Weis, Opt. Commun. {\bf 74}, 301 (1990); E. Pfleghaar, J.
Wurster, S. I. Kanorsky, and A. Weis, {\it ibid} {\bf 99}, 303
(1993); A. J. Wary, D. J. Heading, and J-P Connerade, J. Phys. B
{\bf 27}, 2229 (1994).

\bibitem{connerade_book}
For extensive discussion on MOR and many interesting applications see:
J-P. Connerade, {\em Highly Excited Atoms}, (Cambridge University Press, 1998).

\bibitem{MOR_sat2}
P. Avan and C. Cohen-Tannoudji, J. Physique Lett. {\bf 36}, L85 (1975);
S. Giraud-Cotton, V. P. Kaftandjian, and L. Klein, Phys. Rev A {\bf 32}, 2211
(1985);
F. Schuller, M. J. D. MacPherson, and D. N. Stacey, Opt. Commun. {\bf 71}, 61
(1989).

\bibitem{MOR_sat1}
K. H. Drake and W. Lange, Opt. Commun. {\bf 66}, 315 (1988);
P. Jungner, T. Fellman, B. Stahlberg, and M. Lindberg, {\it ibid}, {\bf 73},
38 (1989).

\bibitem{gsa-sat}
G. S. Agarwal, P. Anantha Lakshmi, J-P Connerade, and S. West,
J. Phys. B {\bf 30}, 5971 (1997).

\bibitem{gawlik_review}
For a recent review of MOR with laser sources see: W. Gawlik, in
``{\em Modern Non-linear Optics}", ed. M. Evans, and S. Kielich, Advances in
Chemical Physics Series vol. {\rm LXXXV}, Part 3 (Wiley, New York, 1994).

\bibitem{voigt}
M. Yamamoto and S. Murayama, J. Opt. Soc. Am. {\bf 69}, 781 (1979).

\bibitem{MOR_inclined}
F. Schuller, R. B. Warrington, K. P. Zetie, M. J. D. Macpherson, and
D. N. Stacey, Opt. Commn. {\bf 93}, 169 (1992).

\bibitem{cold}
G. Labeyrie, C. Miniatura, and R. Kaiser, arXiv: physics/0103045.

\bibitem{biref_Lshift}
A. M. Bonch-Bruevich, N. N. Kostin, and V. A. Khodovoi,
JETP Lett. {\bf 3}, 279 (1966).

\bibitem{biref_incoher}
W. Harpper, Prog. Quantum Electron. {\bf 1}, 53 (1970).

\bibitem{liao:76}
P. F. Liao and G. C. Bjorklund, Phys. Rev. Lett. {\bf 36}, 584 (1976);
Phys. Rev. A {\bf 15}, 2009 (1977).

\bibitem{hansch:76}
C. Wieman and T. W. H\"{a}nsch, Phys. Rev. Lett. {\bf 36}, 1170 (1976);
R. Teets, R. Feinberg, T. W. H\"{a}nsch and A. L. Schawlow,
Phys. Rev. Lett. {\bf 37}, 683 (1976).

\bibitem{heller:81}
Yu. Heller, V. F. Lukinykh, A. K. Popov, and V. V. Slabko,
Phys. Lett. {\bf 82A}, 4 (1981);
see also S. Cavalieri, M. Matera, F. S. Pavone, J. Zhang, P. Lambropoulos,
and T. Nakajima, Phys. Rev. A {\bf 47}, 4219 (1993).

\bibitem{stahlberg:90}
B. St{\aa}hlberg, P. Jungner, T. Fellman, K.-A. Suominen, and S. Stenholm, Opt.
Commun. {\bf 77}, 147 (1990);
K.-A. Suominen, S. Stenholm, B. St{\aa}hlberg, J. Opt. Soc. America B {\bf 8},
1899 (1991).

\bibitem{RI_scully}
M. O. Scully, Phys. Rev. Lett. {\bf 67}, 1855 (1991).

\bibitem{magneto_theory}
M. O. Scully and M. Fleischhauer, Phys. Rev. Lett. {\bf 69}, 1360 (1992);
Phys. Rev. A {\bf 49}, 1973 (1994);
M. Fleischhauer, A. B. Matsko, and M. O. Scully, Phys. Rev. A {\bf 62}, 013808 (2000).

\bibitem{magneto_expt}
V. A. Sautenkov, M. D. Lukin, C. J. Bednar, I. Novikova, E. Mikhailov,
M. Fleischhauer, V. L. Velichansky, G. R. Welch, and M. O. Scully,
Phys. Rev. A {\bf 62}, 023810 (2000);
I. Novikova, A. B. Matsko, V. A. Sautenkov, V. L. Velichansky,
G. R. Welch, and M. O. Scully, Opt. Lett. {\bf 25}, 1651 (2000).

\bibitem{magneto_budker}
D. Budker, V. Yashchuk, and M. Zolotorev, Phys. Rev. Lett. {\bf 81},
5788 (1998);
D. Budker, D. F. Kimball, S. M. Rochester, and V. V. Yashchuk,
Phys. Rev. Lett. {\bf 85}, 2088 (2000);
D. Budker, D. F. Kimball, S. M. Rochester, V. V. Yashchuk, and M. Zolotorev,
Phys. Rev. A {\bf 62}, 043403 (2000).

\bibitem{budker_slow}
D. Budker, D. F. Kimball, S. M. Rochester, and V. V. Yashchuk,
Phys. Rev. Lett. {\bf 83}, 1767 (1999).

\bibitem{pavon:97}
F. S. Pavone, G. Bianchini, F. S. Cataliotti, T. W. H\"{a}nsch,
M. Inguscio, Opt. Lett. {\bf 22}, 736 (1997).

\bibitem{harris}
S. E. Harris, Phys. Today, Pg. 36, July (1997);
S. E. Harris, G. Y. Yin, M. Jain, H. Xia, and A. J. Merriam, Philos.
Trans. R. Soc. (London) A {\bf 355}, 2291 (1997).

\bibitem{gaeta:98}
S. Wielandy and A. L. Gaeta, Phys. Rev. Lett. {\bf 81}, 3359 (1998).

\bibitem{my_MOR}
A. K. Patnaik and G. S. Agarwal, Opt. Commun. {\bf 179}, 97 (2000).

\bibitem{my_DOP_springer}
A. K. Patnaik and G. S. Agarwal, in {\em Frontiers of Laser Physics and
Quantum Optics}, Eds. Z. Xu, S. Xie, S. -Y. Zhu, M. O. Scully (Springer-Verlag,
Germany), pg. 403.

\bibitem{forston:98}
A. D. Cronin, R. B. Warrington, S. K. Lamoreaux, and E. N. Fortson,
Phys. Rev. Lett. {\bf 80}, 3719 (1998).

\bibitem{fulton}
Y. Chen, C. Lin, and I. A. Yu, Phys. Rev. A {\bf 61}, 053805 (2000);
D. McGloin, M. H. Dunn, and D. J. Fulton, Phys. Rev. A {\bf 62}, 053802 (2001).

\bibitem{eit}
M. Xiao, Y. Li, S. Jin, and J. G. Banacloche, Phys. Rev. Lett. {\bf 74}, 666
(1995);
J. G. Banacloche, Y. Li, S. Jin, and M. Xiao, Phys. Rev. A {\bf 51}, 576 (1995);
Y. Li and M. Xiao, {\em ibid} R2730 (1995).

\bibitem{lwi}
G. Vemuri and G. S. Agarwal, Phys. Rev. A {\bf 53}, 1060 (1996);
G. Vemuri, G. S. Agarwal and B. D. N. Rao, {\em ibid}, 2842 (1996).

\bibitem{demtroder}
W. Demtr\"oder, {\em Laser Spectroscopy} (Springer, Berlin, 1998), Chap.3;
R. Loudon, {\em The Quantum Theory of Light} (Oxford University Press),
Chap.2.

\bibitem{abbr_stegun}
M. Abramowitz and I. A. Stegun, {\em Hand Book of Mathematical Functions},
(Dover Publication, NY, 1972), pg. 279.

\bibitem{kerr}
Note that the paper [E. L. Lago and R. de la Fuente, Phys. Rev. A {\bf 60},
549 (1999)] reports polarization switching in a Kerr medium.


\end{thebibliography}
\end{document}